\documentclass{elsart3p}

\usepackage{natbib}
\usepackage{graphics}
\usepackage{graphicx}
\usepackage{epsfig}

\usepackage{amssymb}

\journal{Nucl. Instrum. Meth. A}
\begin{document}

\begin{frontmatter}



\title{An Intense $\gamma$-ray Beam Line of 10 MeV Order Based on Compton Backscattering}


\author[sinap,gscas]{W. Guo\corauthref{1}},\corauth[1]{Corresponding
author.}\ead{guowei@sinap.ac.cn}
\author[sinap]{W. Xu},
\author[sinap]{J.G. Chen},
\author[sinap]{Y.G. Ma},
\author[sinap]{X.Z. Cai},
\author[sinap]{H.W. Wang},
\author[sinap,gscas]{Y. Xu},
\author[sinap,gscas]{C.B. Wang},
\author[sinap]{G.C. Lu},
\author[sinap]{R.Y. Yuan},
\author[sinap]{J.Q. Xu},
\author[sinap]{Z.Y. Wei},
\author[sinap]{Z. Yan},
\author[sinap]{W.Q. Shen}
\address[sinap]{
Shanghai Institute of Applied Physics, Chinese Academy of
Sciences, Shanghai 201800, China }
\address[gscas]{Graduate School of the Chinese Academy of Sciences, Beijing 100039, China}

\begin{abstract}
Shanghai Laser Electron Gamma Source, a $\gamma$-ray beam line of
10MeV order was proposed recently. The beam line is expected to
generate $\gamma$-ray with maximum energy of 22MeV by backward
Compton scattering between CO$_2$ laser and electron in the 3.5GeV
storage ring of future Shanghai Synchrotron Radiation Facility.
The flux of non-collimated $\gamma$-ray can be $10^9\sim10^{10}$
s$^{-1}$ if a commercial CO$_2$ laser of 100W order output power
is employed and injected with optimized settings.
\end{abstract}


\end{frontmatter}

\section{Introduction}
\label{Intro} Compton backscattering (CBS) of laser light against
relativistic electron could be a promising method to obtain
polarized monochromatic $\gamma$ rays with small divergence angle.
This idea was pointed out by Milburn \cite{MIL63}, Arutyunian and
Tumanian \cite{ARU63} for the first time in 1963.

Among $\gamma$-ray producing techniques, CBS method possesses
several significant advantages \cite{SAL95}. First, Compton
scattering is well understood within QED framework, and there are
definite relations among energy, emitting angle, cross section,
polarization, etc. of the scattered photons. Second, the scattered
photons has a sharp cut-off near the maximum energy and the
largest fraction of photons is in the high energy region. Third,
CBS $\gamma$-rays based on relativistic e$^{-}$ have a very small
divergence angle, that allows experiments with both compact
targets and detectors. Fourth, CBS method provides a convenient
and swift way to steer the polarization of $\gamma$-rays by
changing polarization of injected laser beams.

The first $\gamma$-ray facility based on CBS method, the LADON
beam, started to operate in 1978 \cite{CAS75, MAT77, FED80,
BAB91}. Till now, there are several $\gamma$-ray facilities all
over the world \cite{DAN00}. Most of them produce $\gamma$-ray
from several hundred MeV to several GeV. Also, there are a few MeV
$\gamma$-ray facilities: HI$\gamma$S\cite{PAR01,LIT97} on Duke
storage ring, BL05SS\cite{OHK06,FUJ03} on SPring-8 storage ring
and BL-1\cite{AOK04,LI04} on NewSUBARU.

The future Shanghai Synchrotron Radiation Facility (SSRF)
\cite{ZHA04} provides precious opportunity to build a $\gamma$
source of high quality. Recently, Shanghai Laser Electron Gamma
Source (SLEGS), a $\gamma$-ray beam line based on CBS method has
been proposed. In the coming sections, basic relations of CBS
method will be described briefly, the proposed SLEGS facility and
$\gamma$-ray it generated will be illustrated and compared with
other CBS $\gamma$-ray facilities of ten MeV order.

\section{Basic Relations}

\begin{figure}[htbp]
\includegraphics[width=9cm]{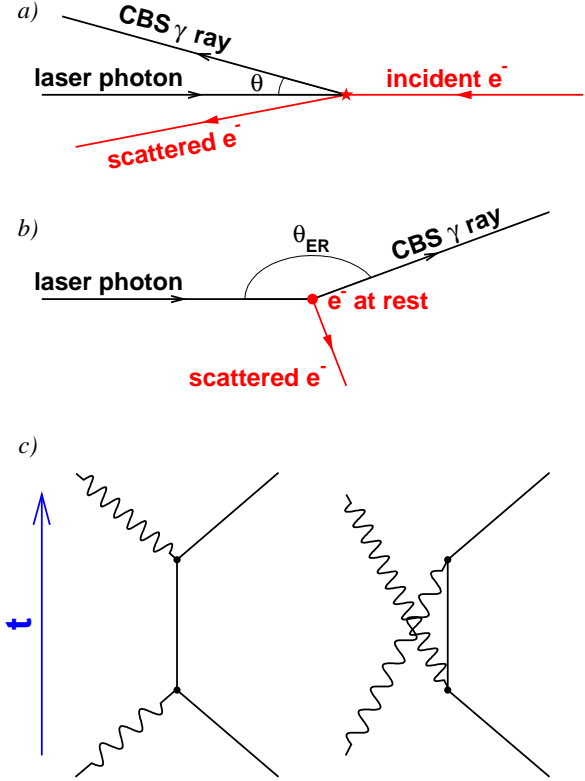}
\caption{Kinematics of CBS process between electron and laser
photon in a$)$ laboratory frame, b$)$ electron rest frame, and
c$)$ Feynman diagram of Compton scattering.} \label{BCS}
\end{figure}
Compton scattering between e$^-$ and photon can be described
within QED framework\cite{GRE94}. The kinematics of CBS in
laboratory (LAB) frame a), electron rest (ER) frame b), and
related Feynman diagram c) is shown as Fig.\ref{BCS}.

Energy of CBS $\gamma$ photon $E_{\gamma}$ is related to the
emitting angle $\theta$:
\begin{equation}
E_{\gamma}=\frac{(1+\beta)E_{L}}
{1-\beta\cos\theta+(1+\cos\theta)E_{L}/E_{0}} \label{ELABTheta}
\end{equation}
where $E_{0}$ is the energy of incident e$^{-}$,
$\gamma=E_{0}/m_{0}c^{2}=1/\sqrt{1-\beta^{2}}$ the relativistic
factor, $m_{0}c^{2}=0.511$ MeV  mass of e$^{-}$ at rest and
$E_{L}$ the energy of incident laser photon. In case of
relativistic e$^{-}$, i.e., $\gamma\gg1$ and $\beta\approx1$,
formula(\ref{ELABTheta}) can be simplified as:
\begin{equation}
E_{\gamma}=\frac{4\gamma^{2}E_{L}}
{1+4\gamma^{2}E_{L}/E_{0}+\gamma^{2}\theta^{2}}
\label{ELABThetaSpl}
\end{equation}
which indicates that the maximum energy of CBS photon is
proportional to $\gamma^{2}$ approximately.

The angular differential cross-section of CBS $\gamma$-rays in ER
frame follows Klein-Nishina formula\cite{KLE29}. The angular
differential cross-section in LAB frame can be derived by Lorentz
transformation. In case of relativistic e$^{-}$, the angular
divergence of CBS $\gamma$-rays in LAB frame is in order of
1/$\gamma$.

When injected laser beam is totally polarized, the degree of
polarization of CBS $\gamma$ photons has a clear relationship with
the scattering angle\cite{DAN00}.

The ideal flux $n_{\gamma}$ of CBS $\gamma$ rays is related to the
impact geometry and spatial densities of electron and laser beams.
The spatial density of electrons is treated as product of Gaussian
distributions, the corresponding widths are determined by beam
qualities and TWISS functions of the storage
ring\cite{ROS94,BUO94,JIN01}. The spatial density of laser is also
treated as product of Gaussian distributions, the corresponding
widths are determined by optics of Gaussian beams\cite{DAV96}.

In a typical head-on setup, the ideal flux of CBS $\gamma$ rays
is:
\begin{equation}
n_{\gamma}=P_0IL
\end{equation}
where $P_0$ is the laser power, $I$ the current of e$^{-}$ beam,
and $L$ the luminosity:
\begin{equation}
L = \frac{2\sigma (1+\beta)}{\pi ceE_L}
\int\frac{ds}{(4\sigma_x^2+w^2)^{\frac{1}{2}}(4\sigma_y^2+w^2)^{\frac{1}{2}}}
\label{EQ:dL-bunch-coaxis}
\end{equation}
where $\sigma$ denotes the total Compton cross section, $w$ the
transversal dimension of laser, $\sigma_{x,y}$ the horizontal and
vertical width of the e$^-$ bunch respectively.

One should notice that the actual luminosity $L^*$ is
significantly affected by geometrical restriction of both injected
laser beam and extracted $\gamma$ beam, this is limited by vacuum
tube dimensions of the storage ring. Henceforth, we applied proper
selection on ideal CBS events during the Mote-Carlo simulation.

\section{SLEGS Facility}

\begin{figure*}[htbp]
\includegraphics[width=15cm]{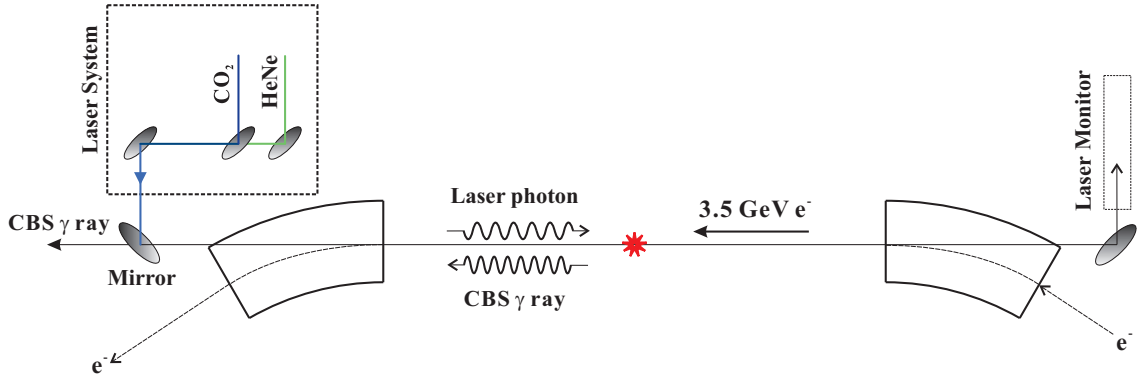}
\caption{A schematic view of future SLEGS facility. }
\label{system}
\end{figure*}
A schematic view of future SLEGS facility is illustrated as
Fig.\ref{system}. Laser beam of operating wavelength is generated
and polarized in the optical chamber, then focused and injected
into SSRF storage ring from front-end downstream to the selected
sector, i.e., the interaction region by a thin mirror. The
operating laser is aligned to the electron beam by a co-axis
visible laser. Once laser beam overlaps e$^{-}$ bunches, Compton
backscattering occurs between relativistic electron and laser
photon, $\gamma$-ray with determined energy and a narrow
divergence angle is generated, transported and then extracted from
the inject front-end. After the rear-end upstream to the
interaction region, remaining laser light is inspected by a set of
monitors which provides feedback and control signals as well.

\subsection{SSRF and its storage ring}
SSRF, now under construction, will operate by year 2009. It will
be a high performance third-generation synchrotron facility
producing high brightness X-ray covering an energy range from
0.1keV to 40keV\cite{ZHA04}. It consists of an 100MeV LINAC, an
100MeV to 3.5GeV booster, a 3.5GeV storage ring and associated
experimental stations. The storage ring has 4 identical long
straight sectors, 16 identical standard straight sectors, and 40
bending magnets\cite{LIU06}. Its designed features are listed in
Table.\ref{T:SSRF-Features}.
\begin{table}[pt]
\centering \caption{Designed features of SSRF storage ring.}
\label{T:SSRF-Features}
\begin{tabular}{@{}r  c  c  c  l  c  l@{}}
\hline\hline perimeter & & ~ & & \hphantom{0}432\hphantom{0} & & m \\
length of long straight & & & & \hphantom{0}12.0\hphantom{0} & & m \\
length of standard straight & & & & \hphantom{0}6.7\hphantom{0} & & m \\
radio frequency & & & & \hphantom{0}499.65\hphantom{0} & & MHz \\
\hline designed energy & & $E_0$ & & \hphantom{0}3.5\hphantom{0} & & GeV \\
momentum dispersion & & $\sigma_p$ & & \hphantom{0}0.1\hphantom{0} & & \%\\
momentum acceptance & & & & \hphantom{0}3.0\hphantom{0} & & \% \\
transversal emittance & & $\varepsilon_{0}$ & & \hphantom{0}3.9\hphantom{0} & & nm$\cdot$rad\\
coupling factor & &  $\kappa$ & & \hphantom{0}1.0\hphantom{0} & & \% \\
natural beam length & & $\sigma_s$ & & \hphantom{0}4.0\hphantom{0} & & mm \\
beam current of single bunch mode & & $I$ & & \hphantom{0}5.0\hphantom{0} & & mA \\
beam current of multi bunch mode & & $I$ & & \hphantom{0}300\hphantom{0} & & mA \\
beam lifetime & & & & \hphantom{0}$>$10\hphantom{0} & & hrs \\
\hline\hline
\end{tabular}
\end{table}

\begin{figure}[htbp]
\includegraphics[width=14cm]{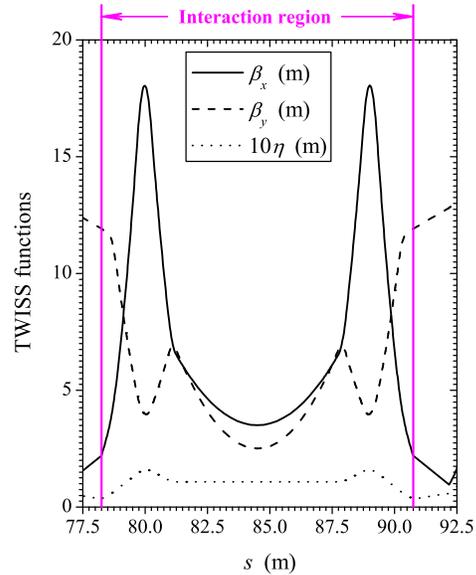} \vspace*{-72pt}
\caption{TWISS functions of SSRF sector 20, calculated by MAD-9.
Straight sector between bending magnets, i.e., the interaction
region, is marked by a pair of vertical lines. } \label{ETWISS}
\end{figure}

The future SLEGS facility can be constructed on one of the
standard straight sectors of SSRF, probably on sector 20. TWISS
functions of sector 20 is shown as Fig.\ref{ETWISS}.

\begin{figure}[htbp]
\includegraphics[width=8cm]{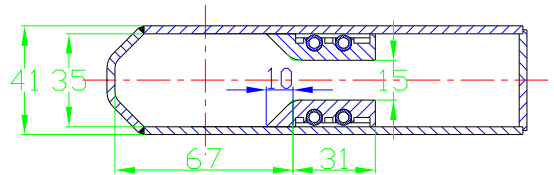} \vspace*{-40pt}
\caption{Sectional dimensions of SSRF vacuum tube for bending
sectors. The dash-dot lines denote the center orbit of electron
bunches. The operating laser beam will be transported through the
slit for outside(right) to inside(left).} \label{TUBE}
\end{figure}

There are two kinds of vacuum tube assembled along SSRF storage
ring. Vacuum tube for the bending sectors has a $\pm7.5$ mm
vertical water-cooling slit to encapsulated microwave radiation of
the RF, shown as Fig.\ref{TUBE} .
Such slits are the most confining parts for laser beam in
transversal directions, henceforth cause a considerable limitation
on the population of injecting photons, and consequentially limit
the total flux of CBS $\gamma$-ray most.

\subsection{CO$_2$ Laser and Corresponding CBS $\gamma$-ray}

The fundamental wavelength of CO$_2$ laser is 10.64$\mu$m, the
corresponding $E_L$ is 0.117eV. The maximum energy of CBS
$\gamma$-ray of CO$_2$ laser with 3.5GeV electron is about 22MeV.
Commercial CO$_2$ laser is widely used in many industrial fields.
It has several advantages compared to molecular gas lasers with
longer wave lengthes:
\begin{itemize}
\item well developed, compact, and with high qualities; \item high
output power that promises a high flux of CBS $\gamma$-ray; \item
run under both continuous and pulsed mode; \item relatively low
cost and low technological risk; \item easy to manage and control.
\end{itemize}

\begin{figure}[htbp]
\includegraphics[width=12cm]{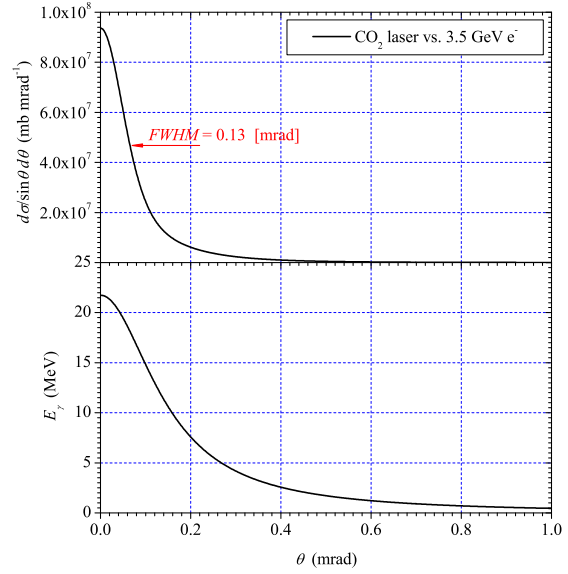}\vspace*{-12pt}
\caption{The angular differential cross section (upper panel) and
energies as a function of emitting angle $\theta$ (lower panel) of
CBS $\gamma$-ray from CO$_2$ laser against 3.5GeV electron. }
\label{CSQLab}
\end{figure}

The angular differential cross section and energies of CBS
$\gamma$-ray from CO$_2$ laser against 3.5GeV electron is shown as
Fig.\ref{CSQLab}.
Since maximum energy of BCS $\gamma$-ray ($\sim$22 MeV) is within
the momentum acceptance of SSRF storage ring($\pm105$ MeV), all
scattered electrons will be re-accelerated. Hence SLEGS can
operate parasitically, independent of other users on SSRF in
principle.

\subsection{Luminosity of CBS $\gamma$-ray}

\begin{figure}[htbp]
\includegraphics[width=9cm]{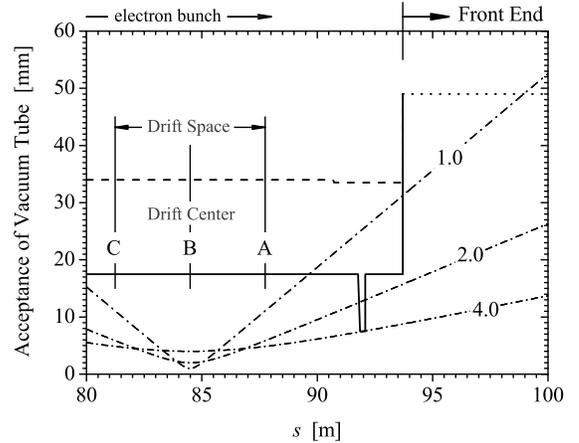}
\caption{Transversal restriction of SLEGS vacuum tube. The
solid(dash) line represents the horizontal(vertical) boundary.
Vacuum tube of the front end of SLEGS beam line is not finally
determined yet, marked as a dotted line. The dash-dot lines
represents transversal dimension $w$ of CO$_2$ laser focusing on
the drift center with different waist widths $w_0=1.0, 2.0$ and
4.0 mm respectively. Direction of electron bunch is shown at the
top. } \label{Acceptance}
\end{figure}

The transversal restriction of SLEGS vacuum tube to the CO$_2$
laser beam is illustrated as Fig.\ref{Acceptance}. As we have
mentioned above, slit of the bending magnet is the most confining
position for injected CO$_2$ laser.

We have developed a C++ program based on Monte-Carlo method to
simulate the generation of CBS $\gamma$-ray.
First, configuration of SLEGS, laser/electron bunch parameters and
TWISS functions are loaded; then, the related ideal luminosity
with infinite acceptance is calculated. Second, the program starts
to sample space coordinates of ideal CBS events according to the
ideal luminosity distribution. Third, the program samples momentum
coordinates for each incident laser photon, and checks if it can
pass the geometrical restriction or not. Fourth, the scattered
angles of CBS $\gamma$-ray are sampled according to differential
cross section of Compton scattering. Last, the program checks if
the CBS $\gamma$ ray is "actual", i.e., the $\gamma$ ray can pass
the geometrical restriction or not.

\begin{figure}[htbp]
\includegraphics[width=10cm]{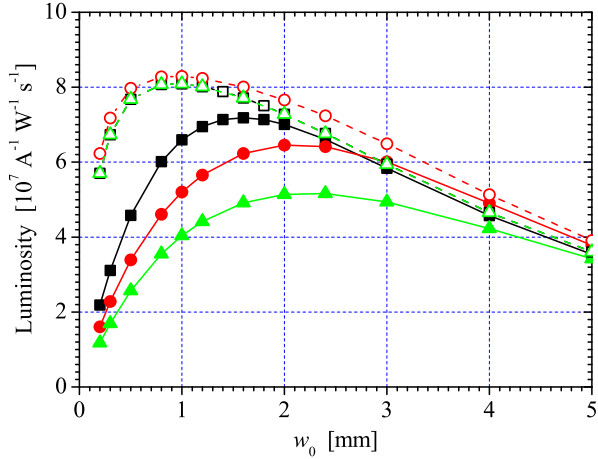}
\caption{Luminosity of CBS $\gamma$ rays in a head-on setup of
CO$_2$ laser against SLEGS electron beam in multi bunch mode. See
the text for explanations. } \label{SIM1}
\end{figure}

In a head-on setup, flux of CBS $\gamma$ rays is related to the
focusing position and waist dimension of laser beam, shown as
Fig.\ref{SIM1}. 
Since the designing of SLEGS front end is now undergoing, we
consider the flange connecting storage ring and front end as the
inspecting plane. The laser power on the inspecting plane is
considered to be the injected power, and CBS $\gamma$ rays that
reach the inspecting plane are considered to be actual. We tried
three different focuses for CO$_2$ laser: at the down-stream end
(A), center (B) and up-stream end (C) of the drift space
respecting to the beam direction of electron bunch, as shown in
Fig.\ref{Acceptance}. In Fig.\ref{SIM1}, solid and open symbols
represent actual($L^*$) and ideal($L$) values respectively;
squares, circles and triangles represent luminosity of different
laser focusing position at A, B and C respectively. As we can see
in Fig.\ref{SIM1}, although lasers with waist dimension of $\sim$
1mm correspond to maximum $L$, the related values of $L^*$ are
evidently limited by geometrical restriction.

The most optimized waist dimension for CO$_2$ laser is around
1.6mm, while the focus is on the down-stream side. In that case,
we can obtain a maximum $L^*\sim7\times10^7$
A$^{-1}$W$^{-1}$s$^{-1}$. Considering SSRF storage ring running
under multi bunch mode($I_m=300$ mA), the flux of BCS $\gamma$-ray
is expected to be $\sim10^{10}$ s$^{-1}$ if a CW CO$_2$ laser of
500 W is employed.

\begin{figure}[htbp]
\includegraphics[width=8cm]{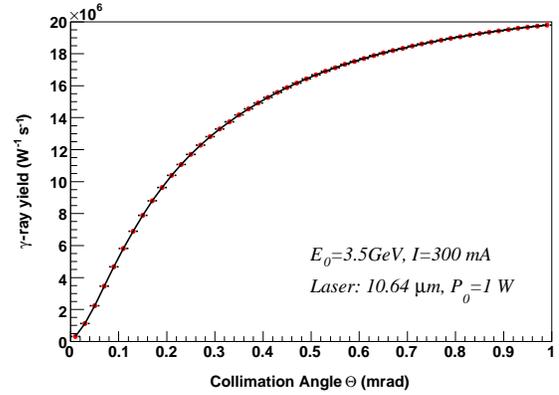}
\caption{Estimated flux of CBS $\gamma$ rays as a function of
collimation angle. } \label{COLL}
\end{figure}

\begin{figure}[htbp]
\includegraphics[width=8cm]{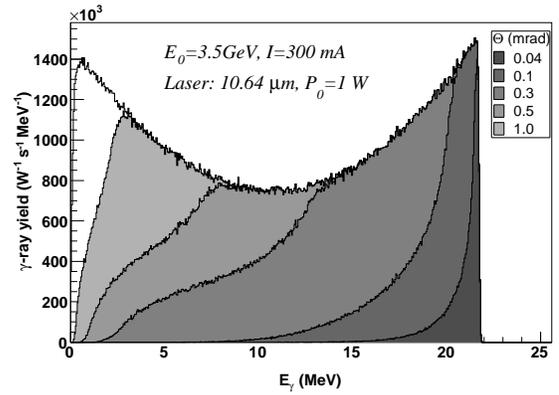}
\caption{The energy spectrum of CBS $\gamma$-ray of different
collimation angle. The histograms filled with gray scale stand for
collimation angle $\Theta$ of 0.04, 0.1, 0.3, 0.5 and 1.0mrad
respectively. The histogram filled with white represents the
non-collimated energy spectrum.} \label{EGCOLL}
\end{figure}

The estimated flux of CBS $\gamma$-ray per unit laser power
gathered within a given collimation angle $\Theta$ is illustrated
as Fig.\ref{COLL}.
The energy spectrum of CBS $\gamma$-ray of different collimation
angle is illustrated as Fig.\ref{EGCOLL}.

\section{Discussion}
Several CBS $\gamma$-ray facilities of 10 MeV order are listed in
Table.\ref{T:Comparation}.
\begin{table*}[pt]
\centering \caption{CBS $\gamma$-ray facilities of 10 MeV order.}
\label{T:Comparation}
\begin{tabular}{@{}r  c  l  c  l  c  l  l  l  l  l  l@{}}
\hline\hline facility & ~ & laser type & ~ & $\lambda$ ($\mu$m)& ~ & $P_0$ (W)& ~ & $E^{\max}_{\gamma}$ (MeV)& ~ & $\gamma$-ray flux (s$^{-1}$)\\
\hline
HI$\gamma$S on Duke storage ring\cite{PAR01} & ~ & free electron laser & &  & &  & & 2$\sim$58 & & $5\times10^7$\\
BL05SS on SPring-8\cite{OHK06} & & CO$_2$ pumped CH$_3$OH laser & & 119 & & 2 & & 10 & & 10$^5$\\
BL-1 on NewSUBARU\cite{AOK04} & & Nd:YVO$_4$ laser & & 1.064 & & 0.74 & & 17.6 & & $1.3\times10^7$\\
\hline
SLEGS on SSRF (present)& & CO$_2$ laser & & 10.64 & & 500 & & 21.7 & & $10^{9}\sim10^{10}$\\
\hline\hline
\end{tabular}
\end{table*}
Compared with those facilities, SLEGS has several unique
properties.

First of all, SLEGS is expected to achieve flux of non-collimated
$\gamma$-ray up to $10^{9}\sim10^{10}$ s$^{-1}$ using a commercial
CW CO$_2$ laser of 100W order output power. The high brightness
$\gamma$-ray will be a powerful probe to meet requirements of many
experimental researches.

Second, since SSRF storage ring has a proper energy to generate
$\gamma$-ray of 10MeV order by CO$_2$ laser, SLEGS does not have
to use a complicated far infrared (FIR) laser system like BL05SS
on SPring-8, so that issues of relatively-low output power,
transmission and focusing for operating laser can be avoided.

Last, HI$\gamma$S possesses unique preponderance of the tunable
$E_\gamma^{max}$ by using a tunable free electron laser (FEL).
However, one of the key issues HI$\gamma$S has to handle is the
synchronization between FEL pulses and target electron bunches.
While SLEGS will avoid such an issue since the operating laser
will run under a CW mode.

To meet the potential requirements of $E_\gamma^{max}$ lower than
22MeV, an oblique injection setup for CO$_2$ laser is within our
scope. If the impact angle between laser and electron beam is
changeable, one can obtain $\gamma$-ray with determined maximum
energy. However, such a setup will definitely cause a decreased
$\gamma$-ray flux, thus method like external laser resonator
\cite{WIL01} is necessary to enforce power of the operating CO$_2$
laser.

\section{Summary}
In this paper, SLEGS, a $\gamma$-ray beam line based on backward
Compton scattering was suggested. SLEGS is expected to achieve a
flux of non-collimated $\gamma$-ray up to $10^9\sim10^{10}$
s$^{-1}$ by employing a CW mode commercial CO$_2$ laser of 100W
order output power. The maximum energy of $\gamma$-ray is about
22MeV.

\section*{Acknowledgements}

This work was supported by the Knowledge Innovation Program of
Chinese Academy of Sciences under Contract No. KJCX2-SW-N13, by
the National Natural Science Foundation of China under Contract
No. 10475108, No. 10505026, and by the hundred talent project of
Shanghai Institute of Applied Physics.





\begin{thebibliography}{}

\bibitem[MIL(1963)]{MIL63}R.H. Milburn, Phys. Rev. Lett. 10 (1963) 75
\bibitem[ARU(1963)]{ARU63}F.R. Arutyunian, V.A. Tumanian, Phys. Lett. 4 (1963) 176
\bibitem[SAL(1995)]{SAL95}E.L. Saldin, {\it et al.}, Nucl. Instr. and Meth. A 362 (1995) 574
\bibitem[CAS(1975)]{CAS75}L. Casano, {\it et al.}, Laser Unconventional Optics J. 55 (1975) 3
\bibitem[MAT(1977)]{MAT77}G. Matone, {\it et al.}, Lect. Notes Phys. 62 (1977) 149
\bibitem[FED(1980)]{FED80}L. Federici, {\it et al.}, Nuovo Cimento B 59 (1980) 247
\bibitem[BAB(1991)]{BAB91}D. Babusci, {\it et al.}, Nucl. Instr. and Meth. A 305 (1991) 19
\bibitem[DAN(2000)]{DAN00}A. D'Angelo, {\it et al.}, Nucl. Instr. and Meth. A 455 (2000) 1, and references therein.
\bibitem[PAR(2001)]{PAR01}S.H. Park, {\it et al.}, Nucl. Instr. and Meth. A 475 (2001) 425
\bibitem[LIT(1997)]{LIT97}V.N. Litvinenko, {\it et al.}, Phys. Rev. Lett. 78 (1997) 4569
\bibitem[OHK(2006)]{OHK06}H. Ohkuma, {\it et al.}, Proceedings of EPAC 2006, Edinburgh, Scotland
\bibitem[FUJ(2003)]{FUJ03}M. Fujiwara, Prog. Part. and Nucl. Phys. 50 (2003) 487
\bibitem[AOK(2004)]{AOK04}K. Aoki, {\it et al.}, Nucl. Instr. and Meth. A 516 (2004) 228
\bibitem[LI(2004)]{LI04}D. Li, {\it et al.}, Nucl. Instr. and Meth. A 528 (2004) 516
\bibitem[ZHA(2004)]{ZHA04}Z.T. Zhao and H.J. Xu, Proceeding of EPAC 2004, Lucerne, Switzerland
%
\bibitem[GRE(1994)]{GRE94}W. Greiner and J. Reinhardt, {\it Quantum Electrondynamics}, Springer-Verlag Berling Heidelberg Press, 1994
\bibitem[KLE(1929)]{KLE29}Klein O. and Nishina Y., Z. Physik 52 (1929) 853
\bibitem[ROS(1994)]{ROS94}J. Rossbach and P. Schm\"{u}ser, {\it Basic Course on Accelerator Optics}, CERN Yellow Report 94-01.v.1
\bibitem[BUO(1994)]{BUO94}J. Buon, {\it Beam Phase Space and Emittance}, CERN Yellow Report 94-01.v.1
\bibitem[JIN(2001)]{JIN01}Y.M. Jin, {\it Electron Storage Ring Physics}, University of Science and Technology of China Press, 2001, in Chinese.
\bibitem[DAV(1996)]{DAV96}C.C. Davis, {\it Lasers and Electro-Optics}, Cambridge University Press, 1996
\bibitem[LIU(2006)]{LIU06}G.M. Liu, {\it et al.}, High Energy Physics and Nuclear Physics 30, Supp-I
(2006) 144
\bibitem{WIL01}I. Will, {\it et al.}, Nucl. Instr. and Meth. A 472 (2001) 79

\end{thebibliography}
\end{document}